# VART: A Tool for the Automatic Detection of Regression Faults[*]

Fabrizio Pastore, Leonardo Mariani
University of Milano-Bicocca, DISCo
{pastore,mariani}@disco.unimib.it

## ABSTRACT

In this paper we present VART, a tool for automatically revealing regression faults missed by regression test suites. Interestingly, VART is not limited to faults causing crashing or exceptions, but can reveal faults that cause the violation of application-specific correctness properties. VART achieves this goal by combining static and dynamic program analysis.

## 1 INTRODUCTION

Software systems are continuously evolving artefacts. They change for many different reasons, including feature extension, program enhancement, and bug fixing. Changing and evolving the software is a tricky task because developers have to simultaneously consider the correctness of the change and its effect on the existing functionalities. Indeed changing the software often results in regression problems, which are faults introduced in functionalities that were supposed to be unaffected by the change [7, 8].

Although changes are tested systematically, regression faults frequently remain silent until observed in the field. This is due to the extensiveness of regression test suites that, although designed to cover both the change and the presumably unchanged functionalities of the system, they may miss important corner cases affected by the change [12].

[*]This work has been partially supported by the H2020 Learn project, which has been funded under the ERC Consolidator Grant 2014 program (ERC Grant Agreement n. 646867).

To cope with the intrinsically limited scope of testing, static program analysis can be exploited to formally verify if a given software (e.g., a modified program) satisfies a set of correctness properties [1, 9, 13]. While static analysis can prove that some properties hold, its practical applicability is limited by the need of specifying the correctness properties that must be checked.

To combine the power of static analysis together with the effectiveness of testing, we designed the *Verification Aided Regression Testing* (VART) technique [12], which exploits *test cases* to both automatically generate the correctness properties that must be checked and identify the properties that have been intentionally violated by the changes, and *static analysis* to both discover properties that provably hold before the change and detect properties that have been unintentionally violated by the change. Empirical results show that the combination of these two solutions can augment the classic regression testing process, which consists of running the available test cases after every change, with a *fully automated verification process* that can detect additional regression problems not revealed by the test cases, including *faults that violate application-specific properties* without producing any crash or exception.

In particular, VART can analyze a change from a *base version* to an *upgraded version* as follows. It first uses dynamic analysis, invariant detection in particular [2], to automatically derive *dynamic program properties* from the information collected while running the test cases on the base version of the software. VART then uses static analysis, model checking in particular [3], to verify if the dynamic program properties provably hold for the base version of the software. The properties that are not proved to hold, either because they are false or they are too hard to prove, are discarded, and only the *true properties* are preserved for the next steps of the analysis.

Since some correctness properties that hold for the base version of the software might be intentionally invalidated by a change, for instance because a software requirement has changed, VART automatically detects the likely *outdated properties* before analyzing the upgraded version of the software by running the test cases designed to verify the change. Any property falsified during the execution of these test cases is likely to be a desirable manifestation of the change (we assume the correctness of the executions produced by passing test cases). For instance, in automotive software, a property $acceleration \geq 0$ is likely violated by any test case that covers a change that introduces the possibility to encode the event of braking as a negative acceleration.

Once the outdated properties have been removed, the resulting set of properties are the *non-regression properties*, that is, the correctness properties that are satisfied by the base version of the software and that should be preserved by the change. VART finally uses static program analysis, again in the form of model checking, to verify the non-regression properties on the upgraded version of the software. Violations, together with counterexamples, are

```
int is_available( t_product* prod ){
        return prod->items > 0;
}
```

**Figure 1: Base Version: function `is_available`.**

reported to the user since they are strong indicators of the presence of regression faults. Note that VART discovers faults on the basis of the violation of application-specific properties derived from the monitored executions. This allows VART to identify faults causing incorrect outcomes, which cannot be detected by program-independent correctness properties, such as the absence of null pointer dereferences.

The VART framework (hereafter VART for simplicity) implements this analysis for C programs, providing: (1) a backend that executes dynamic and static program analyses, (2) a set of command line tools for the automated execution of VART within continuous integration systems, such as Jenkins (https://jenkins.io), and (3) a GUI, implemented as a plug-in for the Eclipse IDE, that provides a set of views for supporting developers in analysis and debugging tasks. A video about the VART tool is available at https://youtu.be/E6eUraMc0x0.

This paper focuses on presenting the framework that implements the VART technique, more precisely it first presents how VART can be used to effectively debug a regression fault (Section 2), then illustrates the components that are part of the tool (Section 3), overviews the empirical results achieved with well known open-source software (Section 4), and provides final remarks (Section 5).

## 2 REVEALING REGRESSION FAULTS

This Section shows how VART can be used to reveal regression faults using a running example.

The program considered in this example implements a simplified library for managing a store. The two software versions differ for the implementation of the function `is_available` that returns an integer value representing whether a product is available or not. The base version of function `is_available`, shown in Figure 1, returns 1 if a product is available, 0 otherwise. The upgraded version of the function, shown in Figure 2-b, in addition to returning 0 and 1, can return −1 for products that are out of catalog.

The upgraded function implements the new requirement to support out of catalog products, but also introduces a regression fault in function `available_products`, shown in Figure 2-b. Function `available_products` receives as input a list of products, and counts the number of available products by summing the values returned by function `is_available`, see the 4th line in function `available_products`. Function `available_products` works properly in the base version of the software, but fails in the upgraded program when processing a list that contains out of stock items. When function `is_available` returns −1, the sum computed in function `available_products` is decreased by one, producing an incorrect result, for example variable `total` may become negative.

Although changes are regularly tested, developers may fail to cover all the relevant cases. For instance, in this case the developer may implement a test case that checks if the upgraded version of function `is_available` returns −1 when the item passed as input is out of stock, omitting to design new test cases with out of stock items for function `available_products`. VART can effectively address these cases revealing the missed regression fault.

### 2.1 Revealing Regression Faults

When VART is executed on the running example, it immediately reports a regression fault in the upgraded version of the software. The only inputs required by our tool are the paths to the project folders containing the base and the upgraded versions of the software and the name of the executables. VART requires the presence of a `Makefile` to run the test cases of the program in the project folder. The inputs are saved in the configuration file `Store.bctmc` stored in an Eclipse project, as shown in Figure 2-a. Once the analysis is started VART executes every analysis step in background.

VART shows the results of the analysis in the Eclipse GUI, reusing standard Eclipse views when possible. VART uses both the *Problems* view (Figure 2-d) and the *Regressions* view (Figure 2-c) to show the list of non-regression properties violated by the code opened in the editor, if any, and by the whole program, respectively. The presence of non-regression properties violated by the upgrade indicates the presence of regressions. In the example, the violation of the property total ≥ 0 in function `available_products` is the effect of the change in function `is_available`, which may return −1 and thus cause the generation of negative values for variable `total`.

VART also annotates the source code editor with information about the identified regression problems. For example, the red marks next to the code visualized in the editor correspond to the program locations that falsified at least a non-regression property (Figure 2-b). The properties corresponding to the marks can be visualized by moving the cursor over the mark, as shown in Figure 4. In this case, the property prods == 0 has been preserved by the upgrade, while the property total ≥ 0 has been falsified.

To let developers to precisely investigate the nature of the regression problems, VART can report the complete list of the dynamic, outdated, valid, and invalid properties (Figure 2-e). The outdated and valid properties can also be visualized contextually in the source code, similarly to the invalid properties, as shown in Figure 5. This information can be useful during debugging. For instance, the fact that the property return == 0 || return == 1 is outdated clearly points the developer at the change that modified the behavior of function `is_avalable`, which is the change responsible for the negative values assigned to variable `total` in function `available_products`.

Finally, VART can visualize the complete list of non regression properties in the view *Non regression properties* (Figure 2-f).

### 2.2 Counter-example Driven Debugging

To further support debugging, VART can generate a counterexample trace that shows how each violated property can be actually violated by an execution of the software. The counterexample trace is visualized in Eclipse using a modified version of the CBMC plug-in [3]. Software engineers can navigate the trace backward, starting from the point of violation of the property, that is, the end of the trace, to identify the variable assignments that caused the property violation. Figure 3 for example shows that the violation of the property total ≥ 0 (Figure 3-a) is caused by the assignment of value −1 to variable `total` (Figure 3-b). The value −1 is returned by function is_available as shown in Figure 3-c. This trace suggests that function available_products should be fixed by adding the support to the case of function is_avalable returning −1.

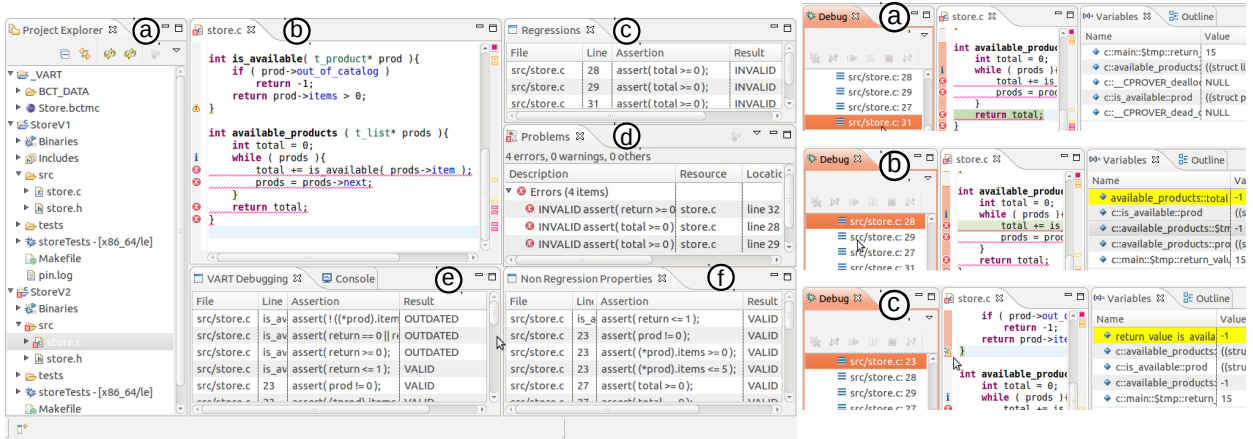

Figure 2: Eclipse with VART views (properties are shown in form of assertions).

Figure 3: Inspection of a counterexample for property total $\geq 0$. The variables with a yellow background are the ones updated in the lines selected (orange) in the left panel.

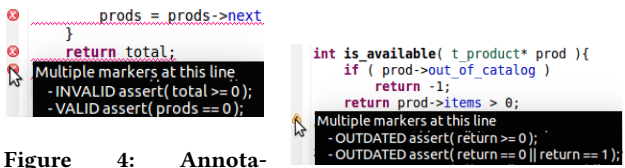

Figure 4: Annotation showing a non-regression property invalid for the upgraded software.

Figure 5: Annotation showing two outdated properties.

## 3 FRAMEWORK OVERVIEW

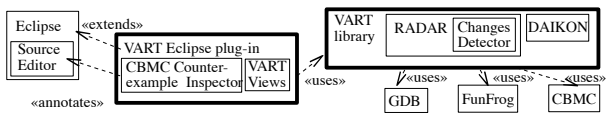

Figure 6: VART components.

This section describes how the components in VART interact to automatically identify regression faults.

VART is implemented in Java according to the architecture shown in Figure 6. The GUI is an Eclipse plug-in (the *VART* Eclipse plug-in) that provides all the views described in Section 2. The functionalities to generate and check program properties are implemented in a standalone JAR library (the *VART library*) that, although released with the plug-in, can be used independently. VART depends on two external tools: the GDB debugger [6], and a bounded model checker, either CBMC [1] or FunFrog [13].

The following paragraphs describe how the VART components cooperate to perform the four main steps of the VART technique: (1)

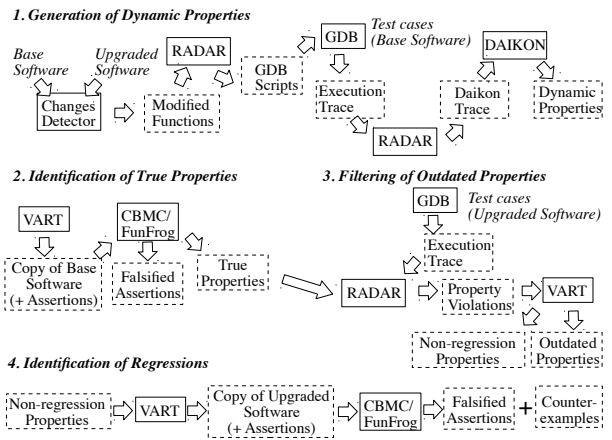

Figure 7: VART workflow.

generate dynamic properties, (2) identify true properties, (3) filter outdated properties, and (4) check non-regression properties. The workflow covering all these four steps is shown in Figure 7.

### 3.1 Generation of Dynamic Properties

VART generates dynamic properties by first collecting the values assigned to the variables in the base version of the software while executing its test cases. The collected values are then passed to the Daikon [2] invariant detector to identify likely assertions on the monitored variables. VART uses RADAR [10, 11] to automatically generate GDB scripts that record variable values. To reduce the cost of the analysis, VART restricts data collection to the functions changed by the upgrade, their callers, and their callees. The value of any variable in the scope of these monitored functions is collected at runtime. Traces are generated in the format processed by Daikon.

Daikon derives properties in the form of Boolean expressions that are satisfied by every occurrence of the variable values recorded in the trace files. In the case of line 28 of file *store.c* for example Daikon generates the property *total* ≥ 0.

## 3.2 Identification of True Properties

The generated dynamic properties may overfit the values observed during the execution of the test cases, which means that they may not hold for every execution of the software. For instance, in case the highest value returned by function `available_products` is 25, Daikon may generate the property *return* ≤ 25, which is a property overfitting the observations. These properties may produce false alarms that may annoy the user of the tool. VART eliminates overfitting and inaccurate properties using static program analysis, thus keeping for the next phases of the analysis only the *true properties*, which are properties that provably hold for every possible execution of the software.

To identify true properties, VART uses bounded model checking technology. More specifically VART integrates CBMC [3] and Fun-Frog [13], two bounded model checkers that can be executed on the source code annotated with assertions to determine the true and false assertions. To run bounded model checking, VART automatically generates a copy of the software with the dynamic properties annotated as assertions. Since CBMC and FunFrog may have different performance depending on the program under analysis, the choice among them is left to the developer.

Bounded model checking may experience some scalability issues when analysing large systems. To mitigate this issue, we scoped the analysis by limiting to N (default is 5 in VART) the number of times loops are unrolled. We found this configuration a good compromise between soundness and scalability, as shown by the fact that although in principle the scoped analysis may fail to discard false properties, it never had an impact on the results reported to users in our experiments. In addition, VART runs the analysis using the callers of the modified functions as entry point, instead of using the main entry point of the software. This choice may discard some properties that are actually true when considering the whole program, but has the benefit to not keep any false property, which may cause false alarms at a later stage of the analysis, and make the analysis feasible even for large systems.

## 3.3 Filtering of Outdated Properties

The true properties identified by VART characterize the behavior of the base version of the software. However the changes implemented in the software may intentionally invalidate some of these properties. We call them *outdated properties*.

VART can automatically identify and drop outdated properties. To achieve this result, VART monitors the execution of the test cases designed to test the change in the upgraded version of the software using Radar [10, 11]. The values collected dynamically during the execution of the test cases are checked with the true properties derived for the base version of the software and any violated property is labeled as outdated and dropped from the analysis. The remaining properties, the *non-regression properties*, are the ones that must be preserved by the update.

## 3.4 Identification of Regressions

To identify regression faults, VART uses again bounded model checking. To this end VART first generates a copy of the upgraded software inserting all the non-regression properties into the code as program assertions. The program is analyzed with a bounded model checker, and the property violations are reported to the user as strong indicators of the presence of a regression fault.

The counterexamples produced by model checking are also presented to the user in a navigable form to ease debugging. Compared to the counterexamples natively produced by the model checkers, VART adjusts line numbers so that the statements in the counterexamples can be mapped to the program without the assertions embedded in the code.

## 4 EMPIRICAL RESULTS

VART has been evaluated in terms of its capability to discover regression faults, its ability to provide sound results despite the usage of incomplete test suites, and its precision in the generation of the alarms [12].

To evaluate its effectiveness, we used VART to analyze changes in two popular open-source string manipulation systems: Sort [4] and Grep [5]. This study resulted in the identification of two regression faults that were not revealed by the projects test suites. The presence of these two faults is confirmed by the entries in the bug repositories of the two projects.

To investigate how sensitive to the completeness of the test suite VART is, we analyzed several versions of GREP using incrementally smaller test suites. Results show that VART can successfully identify regression faults even in the presence of weak test suites. For example, VART has been able to identify regression faults using test suites only covering 25% of the branches in Grep.

Finally, in all the experiments, we observed that the combination of dynamic program analysis and model checking as defined in VART successfully discarded the false (overfitting) and outdated properties. In particular, VART reported only actual faults, that is, non-regression properties violated by faults in the code, and never reported false alarms, that is, properties intentionally violated by changes that affect the behaviour of the software.

## 5 CONCLUSION

VART is a technique that combines static and dynamic analysis to automatically discover regression faults that violate application-specific program properties. VART has been effective with several subtle regression faults [12].

This paper presents the VART tool, which provides a library to execute the VART analysis in a batch fashion and an environment to analyze software upgrades interactively. VART is implemented as an Eclipse plug-in and provides several views to inspect the program properties that can be generated at the various stages of the analysis, to visualize the detected problems contextually in the source code, and to navigate the counterexamples.

VART can be downloaded from the following URL http://www.lta.disco.unimib.it/tools/vart/.